\documentclass[aps,pra,onecolumn,nofootinbib,notitlepage,11pt,tightenlines]{revtex4-1}
\usepackage{amsmath,amssymb,amstext,braket, graphicx,color,xfrac,accents,tabulary,booktabs,xcolor,subcaption,cases}
\usepackage[unicode=true,bookmarks=true,
bookmarksnumbered=false,bookmarksopen=false,
breaklinks=false,
pdfborder={0 0 1},
backref=false,colorlinks=true]{hyperref}

\begin{document}
\baselineskip=14pt
\hfill CALT-TH-2020-012
\hfill

\vspace{0.5cm}
\thispagestyle{empty}

\title{Quantum Space, Quantum Time, and Relativistic Quantum Mechanics}

\author{Ashmeet Singh}
\email{ashmeet@caltech.edu}
\affiliation{Walter Burke Institute for Theoretical Physics, California Institute of Technology, Pasadena, CA 91125}

\newtheorem{thm}{Theorem}[section]
\newtheorem{prop}[thm]{Proposition}
\newtheorem{lem}[thm]{Lemma}
\newtheorem{cor}[thm]{Corollary}
\newtheorem{defn}[thm]{Definition}
\newtheorem{prob}[thm]{Problem}

\newtheorem{example}[thm]{Example}
\newenvironment{eg}{\begin{example}  \rm  }{    \end{example} }

\newcommand{\pf}{\noindent {\bf Proof: }}
\newcommand{\eop}{{\hspace*{\fill}$\square$}}
\newcommand{\eope}{{\hspace*{\fill}\square}}

\newcommand{\defeq}{\vcentcolon=}
\newcommand{\eqdef}{=\vcentcolon}
\newcommand{\pr}{\prime}
\newcommand{\mbf}[1]{\mathbf{#1}}
\newcommand{\mrm}[1]{\mathrm{#1}}
\newcommand{\hs}{\mathcal{H}}

\newcommand{\hst}{\mathcal{H}_{t}}
\newcommand{\hsx}{\mathcal{H}_{\vec{x}}}

\newcommand{\hsspin}{\mathcal{H}_{\mrm{spin}}}
\newcommand{\F}{\mathcal{F}}
\newcommand{\snk}{s^{k}_{n}(x)}
\newcommand{\wnk}{w^{k}_{n}(x)}

\newcommand{\be}{\begin{equation}}
\newcommand{\ee}{\end{equation}}

\newcommand{\spsi}{\ket{\psi}}
\newcommand{\Dim}{\textrm{dim\,}}
\newcommand{\sphi}{\ket{\phi}}
\newcommand{\opphi}{\hat{\phi}}
\newcommand{\oppi}{\hat{\pi}}
\newcommand{\ham}{\hat{H}}
\newcommand{\lcb}{\left[}
\newcommand{\rcb}{\right]}
\newcommand{\Tr}{\mathrm{Tr}}
\newcommand{\bnu}{\hat{b}_{\nu}}
\newcommand{\bnudag}{\hat{b}^{\dag}_{\nu}}
\newcommand{ \Mpt }{M^{2}_{p}}

\newcommand{ \toptr }{\hat{t}}
\newcommand{ \pt }{\hat{p}_t}

\newcommand{ \x }{\hat{x}}
\newcommand{ \X }{\hat{X}}
\newcommand{ \px }{\hat{p}_x}

\newcommand{ \y }{\hat{y}}
\newcommand{ \py }{\hat{p}_y}

\newcommand{ \z }{\hat{z}}
\newcommand{ \pz }{\hat{p}_z}

\newcommand{\eye}{\hat{\mathbb{I}}}
\newcommand{ \J }{\hat{\mathbb{J}}}
\newcommand{\Ult}{\hat{U}}

\newcommand{\kett}[1]{\left| #1 \right\rangle\bigr\rangle}
\newcommand{\braa}[1]{ \left\langle\bigl\langle  #1 \right|}

\newcommand{\matwo}[4]{\left( \begin{array}{cc}#1& #2 \\ #3 &#4 \end{array} \right)}

\newcommand{\dmat}[1]{ {#1}}

\def\({\left(}
\def\){\right)}
\def\[{\left[}
\def\]{\right]}
\def\llangle{\left\langle}
\def\rrangle{\right\rangle}

\newcommand{\eins}{\mbox{$1 \hspace{-1.0mm} {\bf l}$}}


\definecolor{purple}{rgb}{0.5,0,0.5}
\newcommand{\ashmeet}[1]{ { \color{blue} \footnotesize (\textsf{AS}) \textsf{\textsl{#1}} }}




\hypersetup{pdftitle={Quantum Space, Quantum Time, and Relativistic Quantum Mechanics}, pdfauthor={Ashmeet Singh}, citecolor=blue,linkcolor=blue,urlcolor=blue,citecolor=blue}

\begin{abstract}
We treat space and time as bona fide quantum degrees of freedom on an equal footing in Hilbert space. Motivated by considerations in quantum gravity, we focus on a paradigm dealing with linear, first-order Hamiltonian and momentum constraints that lead to emergent features of temporal and spatial translations. Unlike the conventional treatment, we show that Klein-Gordon and Dirac equations in relativistic quantum mechanics can be unified in our paradigm by applying relativistic dispersion relations to eigenvalues rather than treating them as operator-valued equations. With time and space being treated on an equal footing in Hilbert space, we show symmetry transformations to be implemented by unitary basis changes in Hilbert space, giving them a stronger quantum mechanical footing. Global symmetries, such as Lorentz transformations, modify the decomposition of Hilbert space; and local symmetries, such as $U(1)$ gauge symmetry are diagonal in coordinate basis and do not alter the decomposition of Hilbert space. We briefly discuss extensions of this paradigm to quantum field theory and quantum gravity.

\end{abstract}

\maketitle
\tableofcontents

\newpage

\section{Introduction} \label{sec:intro}
Quantum mechanics, in the conventional paradigm, treats time and space on a vastly different footing. Time enters the Schr\"odinger equation as an external, classical parameter that flows independent of the quantum mechanical system. Space, on the other hand, is often elevated to have a genuine quantum mechanical status with corresponding position/momentum operators, unitary transformations, eigenstates, etc. While such an approach works well, both conceptually and mathematically in non-relativistic physics, one would like to treat time and space on an equal footing in relativistic quantum mechanics. Attempts to promote time to an operator have been conventionally argued against, citing the Stone-von Neumann theorem \cite{Kronz2005} due to which the Hamiltonian (with a spectrum bounded from below) and time cannot be bona fide conjugate observables. One then goes on to work with quantum field theory, where time and space are treated as mere labels on a background manifold, and matter is treated quantum mechanically living on this background spacetime.  
\\
\\
In relativistic quantum mechanics of particles\footnote{Often referred to as first quantization, though we will refrain from using this terminology in this paper.}, while one tries to use relativistic dispersion relations to treat time and space symmetrically from an algebraic perspective, their quantum nature is still treated vastly differently as is evident in the Schr\"odinger equation. Efforts to use first order expressions in both the Hamiltonian and momentum lead to rather disparate approaches dealing with Klein-Gordon (spin-$0$ particles) and Dirac (spin-$1/2$ particles) equations. In the case of Klein-Gordon equation, one typically promotes the relativistic dispersion relation $E = \sqrt{|\vec{p}|^{2} + m^{2}}$ for a particle to an operator-valued equation, $\ham = \sqrt{|\hat{\vec{p}}|^{2} + m^{2}}$ to use in the Schr\"odinger equation $\ham \ket{\psi(t)} = i\partial_{t} \ket{\psi(t)}$. Expanding the ``square root" Hamiltonian operator gives a series in all positive even powers of momentum which makes it far from being first-order and leads to issues with non-locality, probabilistic interpretation of the wavefunction etc., as can be found in detail in many advanced undergraduate level texts \cite{bjorken1965relativistic,shankar2012principles,de2007introduction} dealing with relativistic quantum mechanics. The Dirac equation, on the other hand, is able to circumvent this issue by explicitly involving the spin of the particle that makes the Hamiltonian first-order in momentum by using the spinor gamma matrices of the Clifford algebra.  In addition to this, the status of Lorentz transformations from a Hilbert space perspective is often left somewhat ambiguous since time and space are treated rather differently, and hence symmetry transformations are implemented at a classical level. Can these disparities be traced back to, and remedied, by better understanding the quantum mechanical status of space and time? In this paper, we answer this question in the affirmative by treating time and space on an equal footing in Hilbert space.
\\
\\
While quantum theory, in particular its formulation as quantum field theory has been spectacularly successful in predicting outcomes for scattering experiments etc., our primary motivation here is to better understanding aspects of quantum gravity where it is important to examine the status of space and time in the context of quantum mechanics. Approaches to quantum gravity are often plagued by the ``problem of time" \cite{Isham:1992ms, anderson_problem_time, kuchavr1992proceedings,deriglazov2011reparametrization}: reconciling how time enters as an independent, absolute, classical parameter in quantum mechanics; whereas in relativity, time has a relative connotation depending on the observer and distribution of mass-energy. More generally, theories with reparametrization invariance \cite{deriglazov2000local, deriglazov2009improved, deriglazov1995notes} such as classical general relativity, which has general (local) coordinate invariance have a formulation in terms of \emph{constraints} \cite{dirac1950generalized, dirac1988lectures}. Physical states in the theory are the ones that are annihilated by these constraints, and therefore represent the kernel of the constraint operators. In general relativity, Hamiltonian and momentum constraints demand the total energy and momentum to be zero and this is used to identify the physical states. In particular, this is reflected in the Wheeler-DeWitt \cite{dewitt1967quantum} equation, which represents a Hamiltonian constraint of the form $\ham\ket{\Psi} = 0$. In this setup, physical states of the theory do not evolve with respect to an external time. Time evolution, in such a setup would therefore be an emergent feature.
\\
\\
The author and collaborators have argued for a ``quantum-first" approach \cite{Carroll:2018rhc} to quantum gravity where we begin with minimal elements in quantum mechanics in Hilbert space and from it, derive higher-level structures such as space, locality, matter, and eventually, gravity. Similar quantum-first approaches have been advocated by other authors too \cite{giddings2018quantum,giddings2019quantum}. With this motivation in mind, we take a small step in this paper toward understanding the quantum mechanical status of space and time in the context of relativistic quantum mechanics, in a way to lay out groundwork to treat quantum field theory and eventually, gravity in a similar paradigm, appropriately extended. We will treat space and time on an equal footing in Hilbert space, and to this end, we will work with a Hilbert space decomposition of the form,
\begin{equation}
\hs \simeq \hst \otimes \hsx \otimes \hsspin \: ,
\end{equation}
where we have a factor of Hilbert space, $\hst$ for a temporal degree of freedom, a factor, $\hsx$ for a spatial degree of freedom and $\hsspin$ corresponding to the spinorial degree of freedom of the particle. 
Global states in this Hilbert space will neither evolve relative to an external time nor will they have a notion of spatial translations relative to any external coordinate system. With both a Hamiltonian constraint $\J_H$, and momentum constraints $\J_{\vec{P}}$ as central structures in this construction, spatial and temporal translations will be emergent features for physical states. The constraints will be explicitly linear and first order in conjugate momenta, and their joint kernel will define the set of physical states \textit{i.e.} the ones annihilated by the constraints. For such physical states, spatial and temporal translation features emerge as a consequence of entanglement and correlations between different factors of the global Hilbert space $\hs$. Since we are working with a relativistic setup, the compatibility $\lcb \hat{H}, \hat{P} \rcb = 0$ of the Hamiltonian and momentum, two of the generators of the Poincar\'e group will restrict the kinds of theories we can write down and we will discuss this point in some detail. Using these first-order constraints, we will show that Klein-Gordon and Dirac equations in relativistic quantum mechanics can be treated with a uniform approach. This will be done by applying dispersion relations to eigenvalues which appear in constraints, and not treat them as operator-valued equations. With such an approach, the ``square root" Hamiltonian in Klein-Gordon theory is handled naturally at par with the Dirac equation. Along the way, we will also discuss  differences of our setup compared to the usual constructions and how one can attempt to bridge the gap. Treating time and space on an equal footing in Hilbert space, we show symmetry transformations to be implemented by unitary basis changes in Hilbert space. Global symmetries, such as Lorentz transformations, modify the decomposition of Hilbert space; and local symmetries, such as $U(1)$ gauge symmetry are diagonal in coordinate basis and do not alter the decomposition of Hilbert space. 
\\
\\
Our focus in this paper is to simplistically evaluate the quantum mechanical status of space and time with an eye toward relativistic quantum mechanics: recasting the basics of quantum mechanics in a way that can be made amenable to the study of quantum gravity from first principles. The paper is organized as follows. In section \ref{sec:time-space-spin-HS}, we first introduce the concept of internal time treated as a quantum degree of freedom (\`a la Page-Wootters) and then use it as our motivation to lay out the Hilbert space structure treating space and time on an equal footing. Once we have our vector spaces in place, we will then go on to talk about Hamiltonian and momentum constraints in section \ref{sec:constraints} and apply it to Klein-Gordon and the Dirac equations. 
In section \ref{sec:basis_changes}, we will use the power of our Hilbert space construction to identify the status of symmetry transformations in relativistic quantum mechanics as implementing basis changes in Hilbert space. Lorentz transformations will be seen as global changes of factorization of Hilbert space and $U(1)$ gauge symmetry will be enacted as a local basis change while treating time and space symmetrically. We will conclude in section \ref{sec:discussion}, discussing implications and extensions of our construction to quantum field theory and quantum gravity. 

\section{Time, Space and Spin in Hilbert Space} \label{sec:time-space-spin-HS}
\subsection{Inspiration: Time as an Internal Quantum Degree of Freedom}\label{subsec:PW}
We begin by reviewing (an extension of) the Page-Wootters construction \cite{page1983evolution,wootters1984time,page1989itp} which is one of the most famous and elegant approaches to emergent time in quantum mechanics. We will closely follow the work in Ref.  \cite{giovannetti2015quantum} for this quick review. This will serve as motivation for us to generalize its features to treat space and time on an equal footing in Hilbert space in the context of the quantum mechanics of a relativistic particle. The Page-Wootters formulation is one of internal time, where time is treated as an internal quantum degree of freedom and not as an external classical parameter. The global quantum state is static and the apparent ``flow" of time is due to the entanglement and correlations between the temporal degree of freedom with the rest of Hilbert space. \\
\\
The global Hilbert space $\hs$ is factorized into a temporal degree of freedom $\hst$, often called as the ``clock", and the system $\hs_{S}$(what we typically describe in conventional quantum mechanics) (we have used $\simeq$ to denote Hilbert space isomorphisms throughout the paper),
\begin{equation}
\label{eq:PW_HS}
\hs \: \simeq \: \hst \: \otimes \: \hs_{S} \: .
\end{equation}
As we will see, correlations between $\hst$ and $\hs_S$ will lead to effective time evolution for states in $\hs_{S}$ governed by a Hamiltonian. The temporal Hilbert space $\hst$ is taken isomorphic to $\mathbb{L}_{2}(\mathbb{R})$ (akin to the Hilbert space of a single particle on a 1D line in conventional non-relativistic quantum mechanics) and on the space of linear operators on this space $\mathcal{L}(\hst)$, we associate conjugate variables: the ``time" $\toptr$ and its conjugate momentum $\pt$ that satisfy Heisenberg canonical commutation relation (CCR), in units with $\hbar = 1$,
\begin{equation}
\lcb \toptr , \pt \rcb = i \: .
\end{equation} 
A-priori, the conjugate momentum $\pt$ to the time operator $\toptr$ should not be tied in any way to the Hamiltonian. At this stage, we have just specified a standard pair of conjugate operators on the Hilbert space $\hst$. Eigenstates of the time operator $\toptr$ are defined by $\toptr \ket{t} = t \ket{t} \: \forall \: t \in \mathbb{R}$ and these eigenstates follow Dirac orthonormality $\braket{t'|t} = \delta(t -t')$. The Page-Wootters internal time construction then can be written in terms of a constraint operator $\J$ in the linear space of operators $\mathcal{L}(\hs)$,
\begin{equation}
\label{eq:PW_J}
\J \: = \: \pt \otimes \eye_{S} + \eye_{t} \otimes \ham_{S} \: ,
\end{equation}
where $\eye_{t}$ and $\eye_{S}$ are identity operators on $\hst$ and $\hs_{S}$, respectively, and $\ham_S$ is the conventional Hamiltonian for the system. Physical states $\kett{\Psi}$ in the global Hilbert space $\hs$ are identified to be the ones annihilated by the constraint operator $\J$,
\begin{equation}
\J \: \approx \: 0 \: \implies \: \J \: \kett{\Psi}\: = \: 0 \: .
\end{equation} 
We use the double-ket notation $\kett{\Psi}$ to stress the fact that the state is defined on the global Hilbert space $\hst\otimes \hs_S$. Such a technique of quantization based on constraints can be attributed to Dirac \cite{dirac1950generalized,dirac1988lectures}  and also represents the constraint feature of the Wheeler-DeWitt equation. These physical states, which are eigenstates of the constraint operator $\J$ with eigenvalue zero are globally static but encode an apparent flow of time from the perspective of $\hs_S$. Such physical states annihilated by a global constraint are therefore consistent with the Wheeler-DeWitt equation. Conventional time-dependent states of the system are obtained by conditioning the global, physical state $\kett{\Psi}$ with the eigenvector $\ket{t}$ of the time operator $\toptr$,
\begin{equation}
\ket{\psi(t)} \: = \:\big\langle t \kett{\Psi} \: \in \hs_S \: ,
\end{equation}
which obeys the conventional Schr\"odinger equation governed by the Hamiltonian $\ham_{S}$,
\begin{equation}
\label{eq:PW_Schrodinger}
\bra{t}\J\kett{\Psi} \: = \: \bra{t}\pt\otimes\eye_{S} \kett{\Psi} \: + \: \ham_{S}\ket{\psi(t)} \: = 0 \: .
\end{equation}
Inserting a complete set of states on $\hst$ given by $\int dt \: \ket{t}\bra{t} = \eye_{t}$, and remembering that the matrix elements of the conjugate momenta are $\braket{t | \pt | t'} = -i \frac{\partial}{\partial t} \delta(t - t')$, we get the time evolution equation for states $\ket{\psi(t)}$ of the system,
\begin{equation}
\ham_{S} \ket{\psi(t)} \: = \: i \frac{\partial}{\partial t} \ket{\psi(t)} \: ,
\end{equation}
which is indeed the Schr\"odinger equation for $\ket{\psi(t)} \in \hs_S$. Thus we see that effective time evolution for states in the subfactor $\hs_S$ of the global Hilbert space, governed by a Hamiltonian $\ham_S$ can be recovered from a constraint operator. 
\\
\\
Such a construction is succinct and elegant since it gives a strong quantum mechanical notion of a temporal degree of freedom in Hilbert space. It also overcomes Dirac's criticism on treating time as an operator: following the Schr\"odinger equation, one might wish to establish a conjugate relationship between the Hamiltonian and the time operator as canonically conjugate variables but this is not allowed due to the Stone-von Neumann theorem. The theorem demands a set of conjugate operators satisfying the Heisenberg CCR to have their eigenvalue spectra unbounded from below; but for physical theories, the Hamiltonian has a ground state with an energy bounded from below. In the Page-Wootters construction, the time operator $\toptr$ and the system Hamiltonian $\ham_S$ are not conjugates since they act on different Hilbert spaces. There is a bona fide pair of conjugate operators on $\hst$, the time operator $\toptr$ and its conjugate $\pt$ which satisfy the Heisenberg CCR. It's only for the physical states which are annihilated by the constraint $\J$ that leads to a Schr\"odinger evolution for these states. 
\\
\\
While elegant and succinct, the Page-Wootters formulation of internal time treats time as a special, distinguished variable on a vastly different footing than space as is evident from the construction. The nature of the system Hilbert space $\hs_S$ is left open-ended on purpose and has the potential of representing a variety of degrees of freedom, or combinations thereof, including but not limited to space and spin. While one might choose a position or momentum basis for $\hs_S$ in certain examples, it does not have any explicit and universal connection to a spatial degree of freedom as is evident by the lack of a corresponding momentum constraint (just like we have a Hamiltonian constraint associated with the temporal degree of freedom). It therefore, in its current form, is not very amenable to understanding relativistic quantum mechanics and the status of symmetry transformations such as Lorentz transformations etc. in Hilbert space. In an effort to formulate a quantum-first approach to quantum gravity, we would like to be able to treat space and time coordinates on an equal footing in a reparametrization invariant way.
Motivated by the Page-Wootters construction, we now move on to developing the basic framework to treat time and space on an equal footing in Hilbert space with a corresponding Hamiltonian and momentum constraints. The interested reader who would like to delve more into the problem of time in quantum gravity more broadly, the Page-Wootters mechanism, conditional probability approach to time, and allied topics is referred to  \cite{gambini2004loss,gambini2004relational,gambini2009conditional,leon2017pauli,mendes2019time} (and references therein).
\subsection{Hilbert Space Structure}
Our focus in this paper is the quantum mechanics of a relativistic particle in a 3+1 d spacetime, \textit{i.e.} three spatial dimensions and one temporal dimension. Let us begin by introducing the formal Hilbert space structure of the theory, as a modification to the Hilbert space decomposition of Eq. (\ref{eq:PW_HS}) in the Page-Wootters construction above. For a similar Hilbert space construction applied to relativistic ideas of time dilation, please see Ref.  \cite{smith2019relativistic}. Instead of having a system Hilbert space $\hs_S$, we will treat space on an equal footing with time by assigning it as a factor in the Hilbert space decomposition in addition to accounting for the spin of the particle, 
\begin{equation}
\label{eq:HS_decomposition}
\hs \simeq \hst \otimes \hsx \otimes \hsspin \: ,
\end{equation}
where $\hst$ is the Hilbert space associated with a temporal degree of freedom, $\hsx$ with a spatial degree of freedom and $\hsspin$ corresponds to the spinorial degree of freedom. The temporal Hilbert space $\hst$ is again taken isomorphic to $\mathbb{L}_{2}(\mathbb{R})$ (akin to the Hilbert space of a single particle on a 1D line in conventional non-relativistic quantum mechanics) similar to the Page-Wootters case and on the space of linear operators $\mathcal{L}(\hst)$, we associate conjugate variables: the ``time" coordinate operator $\toptr$ and its conjugate momentum $\pt$, which satisfy Heisenberg canonical commutation relation (CCR), in units with $\hbar = 1$,
\begin{equation}
\lcb \toptr , \pt \rcb = i \: .
\end{equation} 
Similar to the Page-Wootters construction, the conjugate momentum $\pt$ to the time operator $\toptr$ a-priori should not be tied to the Hamiltonian in any way. The eigenstates of the time operators are defined in the usual way $\toptr \ket{t} = t \ket{t} \: \forall \: t \in \mathbb{R}$ satisfying Dirac orthonormality $\braket{t|t'} = \delta(t - t')$ and the conjugate momentum $\pt$ generates translations of the $\ket{t}$ eigenstates, $\exp{\(- i \pt \Delta t\)} \ket{t} = \ket{t + \Delta t}$.
\\
\\
For the spatial factors of Hilbert space, since we are working in 3 spatial dimensions, we associate a factor isomorphic to $\mathbb{L}_{2}(\mathbb{R})$ for each orthogonal direction which we choose to label by Cartesian directions $x, y $ and $z$,
\begin{equation}
\hsx \simeq \hs_{x} \otimes \hs_{y} \otimes  \hs_{z} \: ,
\end{equation}
and for each of these factors, we associate conjugate variables satisfying Heisenberg CCR,
\begin{equation}
\lcb \hat{j} , \hat{p}_{j} \rcb = i , \: \: \mrm{for} \:  j = x,y,z \: . 
\end{equation}
Throughout the paper, we use Latin index $j$ to run over the spatial coordinates $j = x,y,x$ and Greek indices $\mu,\nu = 0,1,2,3$ to run over spacetime coordinates. We also define 4-operators $\hat{X^{\mu}}$ and $\hat{P}_{\mu}$ for $\mu = 0,1,2,3$ living in $\mathcal{L}(\hs)$, the set of linear operators on the full Hilbert space, akin to 4-vectors in special relativity in anticipation of making the formulation covariant,
\begin{equation}
\label{eq:X_mu}
\hat{X}^{\mu} \equiv \left( \X^{0} , \X^{1}, \X^{2}, \X^{3}  \right) \: \in \mathcal{L}(\hst \otimes \hsx) \:, 
\end{equation}
where $\X^{0}$ is to be interpreted at $\X^{0} = \toptr \otimes \eye_{\vec{x}} \:$, $\X^{1} = \eye_{t}\otimes \x \otimes \eye_{y} \otimes \eye_{z}$, etc., and similarly its conjugate momentum,
\begin{equation}
\label{eq:P_mu}
\hat{P}_{\mu} = \left( \hat{P}_{0}, \hat{P}_{1}, \hat{P}_{2} , \hat{P}_{3} \right) \in \mathcal{L}(\hst \otimes \hsx) \: ,
\end{equation}
where $\hat{P}_{0} = \pt \otimes \eye_{\vec{x}} \:$, $\hat{P}_{1} = \eye_{t}\otimes \hat{p}_{x} \otimes \eye_{y} \otimes \eye_{z}$, etc. and these conjugate operators satisfy Heisenberg CCR, $\lcb \hat{X}^{\mu},  \hat{P}_{\nu} \rcb = i \delta^{\mu}_{\nu}$, where $\delta^{\mu}_{\nu}$ is the Kronecker delta function.
\\
\\
The spinorial factor of Hilbert space $\hsspin$ will encode information about spin of the particle, and will typically be spanned by the corresponding spinorial matrices. In particular, a spin-0 particle will have $\dim \hsspin = 1$ and a spin-1/2 particle will correspond to $\dim \hsspin = 4$ (as with the spinor gamma matrices in the Dirac equation that describes both the particle and its antiparticle).
\\
We would like to emphasize that in this setup, there is no notion of an external, classical time parameter and consequently, no Schr\"odinger evolution for states (or evolution of operators in the Heisenberg picture). Time is treated on a equal footing with the spatial degree of freedom of a particle and any notion of spatial or temporal translations should be emergent  features as we will see in the next section.
\section{Hamiltonian and Momentum Constraints}\label{sec:constraints}
With our motivation from quantum gravity to deal with theories with reparametrization invariance such as those with coordinate invariance, a characteristic signature of which is a constraint-based formulation. We too would like to formulate our construction in terms of linear, first-order constraints while treating time and space on an equal quantum-mechanical footing. We thus have a formulation in terms of Hamiltonian and momentum constraints, which assert that the total energy and momentum are zero and these constraints identify physical states to be the the ones that are annihilated by them.
\\
\\
At variance with the Page-Wootters construction outlined in section \ref{subsec:PW} that treats time as a distinguished quantum degree of freedom and hence deals only with a Hamiltonian constraint, we are attempting to treat both space and time on an equal footing and this will have us using both Hamiltonian and momentum constraints to identify physical states. Since we are dealing with relativistic quantum mechanics, we require the constraints to commute with each other (since the Hamiltonian and momentum of a relativistic system commute). As a consequence of this setup being applied to relativistic particles (or as some would say, first quantization), we use a collection of constraints parametrized by $\vec{k} \in \mathbb{R}^{3}$ and the corresponding dispersion relation of the particle. We discuss extensions of this construction to field theory in section \ref{sec:discussion} where we can deal with single Hamiltonian and momentum constraints that commute with each other.
\subsection{Klein-Gordon Equation (Spin-0)} \label{sec:KG}
We first analyse the case of relativistic quantum mechanics of a spin-0 particle with rest mass $m$,  which obeys the Klein-Gordon equation. The energy-momentum dispersion relation for the Klein-Gordon free particle is, 
\begin{equation}
E(\vec{k}) = \sqrt{|\vec{k}|^{2} + m^{2}} \: ,
\end{equation}  
for a momentum $\vec{k} \in \mathbb{R}^{3}$ carried by the particle. In the usual construction dealing with the Klein-Gordon equation, one typically promotes the relativistic dispersion relation of the particle to an operator-valued equation, $\ham = \sqrt{|\hat{\vec{p}}|^{2} + m^{2}}$ that is then used in the Schr\"odinger equation $\ham \ket{\Psi} = i\partial_{t} \ket{\Psi}$. Expanding the ``square-root" Hamiltonian operator gives a series in all even powers of momentum that is far from being first-order, and leads to a slew of issues, from non-locality (due to the higher powers of momentum) in the theory, to not having a consistent probabilistic interpretation of the wavefunction. These issues are brought to light in a standard undergraduate-level quantum mechanical text \cite{bjorken1965relativistic,shankar2012principles,de2007introduction} and we do not reproduce these arguments in detail there. Instead, we will recast the physics of the Klein-Gordon equation in the language of linear, first-order Hamiltonian and momentum constraints, and see how it can help us deal with some of these issues.
\\
\\
 The spinorial Hilbert space in this case will have $\dim \hsspin = 1$ since the particle does not have any spin. Hence, each energy/momentum configuration can be labelled by the spatial momentum $\vec{k}$ of the particle. The Hamiltonian constraint of the system, for a given $\vec{k}$, can be written as,
\begin{equation}
\label{eq:KG_JH}
\hat{\mathbb{J}}_{H}(\vec{k}) = \pt \otimes \eye_{\vec{x}} \otimes \eye_{\mrm{spin}} \: + \: \eye_{t} \otimes \eye_{\vec{x}} \otimes \( E(\vec{k}) \: \eye_{\mrm{spin}}\) \: , 
\end{equation}
where we have formally written down the energy $E(\vec{k})$ in the one-dimensional spinorial Hilbert space $\hsspin$ with the identity element $\eye_{\mrm{spin}} = 1$. The momentum constraints, one for each orthogonal direction, parametrized by $\vec{k} \equiv (k_{x}, k_{y}, k_{z})$,
\begin{equation}
\label{eq:KG_JPj}
\hat{\mathbb{J}}_{P_{j}}(\vec{k}) = \eye_{t} \otimes \hat{p}_{j} \otimes \eye_{\mrm{spin}} \: - \: \eye_{t} \otimes \eye_{\vec{x}} \otimes \(k_{j} \: \eye_{\mrm{spin}}\) \: \: \mrm{for} \:  j = x,y,z \: . 
\end{equation}
Physical states $\kett{\psi_{\vec{k}}}$ in Hilbert space $\hs$ are identified to be the ones that are annihilated by the constraints,
\begin{equation}
\label{eq:KG_phys_states_H}
\J_{H}(\vec{k}) \approx 0 \: \implies \:  \left(\pt \otimes \eye_{\vec{x}} \otimes \eye_{\mrm{spin}} \: + \: \eye_{t} \otimes \eye_{\vec{x}} \otimes \( E(\vec{k}) \: \eye_{\mrm{spin}}\) \right) \kett{\psi_{\vec{k}}} = 0 \: , 
\end{equation}
\begin{equation}
\label{eq:KG_phys_states_mom}
\J_{P_{j}}(\vec{k}) \approx 0 \: \implies \: \left(\eye_{t} \otimes \hat{p}_{j} \otimes \eye_{\mrm{spin}} \: - \: \eye_{t} \otimes \eye_{\vec{x}} \otimes \(k_{j} \: \eye_{\mrm{spin}}\)\right) \kett{\psi_{\vec{k}}} = 0 \: \: \mrm{for} \:  j = x,y,z \: . 
\end{equation}
Again, we have used the double-ket notation to explicitly reflect that these states are defined on the full Hilbert space $\hs$. Since the Hamiltonian and momentum constraint operators commute $\lcb \J_{H}, \J_{P_{j}} \rcb = 0$ as one would expect for a relativistic theory (where the generators $\hat{H}$ and $\hat{\vec{P}}$ of the Poincar\'e group commute). We can write down the physical states $\kett{\psi_{\vec{k}}}$ as simultaneous eigenstates of $\J_{H}$ and $\J_{P_{j}}$ with zero eigenvalue (from the constraints Eqs. (\ref{eq:KG_phys_states_H}) and (\ref{eq:KG_phys_states_mom})),
\begin{equation}
\kett{\psi_{\vec{k}}} = \ket{p_{t} = -E(\vec{k})} \otimes \ket{p_{x} = k_{x}} \otimes \ket{p_{y} = k_{y}} \otimes \ket{p_{z} = k_{z}} \: ,
\end{equation}
where $\ket{p_{t} = -E(\vec{k})}$ is the eigenstate of $\pt$ with eigenvalue $\(-E(\vec{k})\)$ and similarly $\ket{p_{j} = k_{j}}$ is an eigenstate of $\hat{p}_{j}$ with eigenvalue $k_{j}$. The physical eigenstate $\kett{\psi_{\vec{k}}}$ of the constraints have a tensor product form in the momentum basis since each term in a given constraint operator commutes with each other. Written in the coordinate basis, these physical eigenstates can be expressed as,
\begin{equation}
\label{eq:KG_psi_k_kett}
\kett{\psi_{\vec{k}}} = \frac{1}{\sqrt{2\pi}}\int dt \:  e^{-iE(\vec{k}) t} \: \ket{t} \otimes \frac{1}{(2\pi)^{3/2}} \int d^{3}x \: e^{i \vec{k} \cdot \vec{x}} \: \ket{\vec{x}} \: , 
\end{equation} 
where $\ket{\vec{x}} \equiv \ket{x}\otimes\ket{y}\otimes\ket{z}$. We can now obtain the wavefunction of the particle by conditioning a global physical state $\kett{\Psi} \in \hs$ on a tensor product basis element $\ket{t} \otimes \ket{\vec{x}}$. The conditioned state lives in $\hsspin$, the spinorial Hilbert space and since in this case, the particle is spinless, $\hsspin$ is one-dimensional, we get the wavefunction of the particle,
\begin{equation}
\Psi(t,\vec{x}) \equiv \bigg(\bra{t} \otimes \bra{\vec{x}}\bigg) \kett{\Psi} \: .
\end{equation} 
We can similarly condition the constraint equations, Eqs. (\ref{eq:KG_phys_states_H}) and (\ref{eq:KG_phys_states_mom}) and get governing evolution equations in space and time for the wavefunction. Let us first do this for the Hamiltonian constraint,
\begin{equation}
\bigg(\bra{t} \pt  \otimes \bra{x} \bigg) \kett{\psi_{\vec{k}}} + \left(\sqrt{|\vec{k}|^{2} + m^{2}} \right) \: \psi_{\vec{k}}(t,\vec{x}) = 0 \: ,
\end{equation}

Inserting a complete set of states on $\hst$ given by $\int dt \: \ket{t}\bra{t} = \eye_{t}$ and remembering that the matrix elements of the conjugate momenta go as $\braket{t | \pt | t'} = -i \frac{\partial}{\partial t} \delta(t - t')$, we get the time evolution equation for the wavefunction $\psi_{\vec{k}}(t,\vec{x})$ corresponding to the physical state $\kett{\psi_{\vec{k}}}$,
\begin{equation}\label{eq:KG_wavefunction_t}
i \frac{\partial}{\partial t} \psi_{\vec{k}}(t,\vec{x}) \: = \:  \left(\sqrt{|\vec{k}|^{2} + m^{2}} \right) \: \psi_{\vec{k}}(t,\vec{x}) \: .
\end{equation}
which represents the analogue of Schr\"odinger equation governing the time evolution of the wavefunction dictated by the ``Hamiltonian", in this case the energy of the particular $\vec{k}$ mode.

One can similarly get an equation that governs the spatial translations of the wavefunction from the momentum constraint of Eq. (\ref{eq:KG_phys_states_mom}),
\begin{equation}\label{eq:KG_wavefunction_x}
-i \vec{\nabla} \psi_{\vec{k}}(t,\vec{x}) \: = \:  \vec{k} \: \psi_{\vec{k}}(t,\vec{x})  \: .
\end{equation}
 The wavefunction solution corresponding to the physical eigenstate $\kett{\psi_{\vec{k}}}$ satisfying Eqs. (\ref{eq:KG_phys_states_H}) and (\ref{eq:KG_phys_states_mom}) can be found from Eq. (\ref{eq:KG_psi_k_kett}),
\begin{equation}
\psi_{\vec{k}}(t,\vec{x}) \: \sim \: \exp{\left(-i E(\vec{k}) t + i \vec{k} \cdot \vec{x}\right)} \: ,
\end{equation}
which, as expected, yields plane wave solutions (we use a $\sim$, and not an exact equality here since individual plane wave solutions are non-normalizable). For completeness, we mention that one can also have a negative sign with $E(\vec{k})$, interpreted as a negative frequency, in the Hamiltonian constraint, $\hat{\mathbb{J}}_{H}(\vec{k}) = \pt \otimes \eye_{\vec{x}} \otimes \eye_{\mrm{spin}} \: +  \: \eye_{t} \otimes \eye_{\vec{x}} \otimes \( -E(\vec{k}) \: \eye_{\mrm{spin}}\)  \approx 0$, and by including this, one can recover both positive and negative frequency solutions of the Klein-Gordon equation. Formally, however, we always keep a positive sign between the $\pt$ term and the frequency term in the Hamiltonian constraint (similar to Eq. \ref{eq:PW_J} in the Page-Wootters construction).
\\
\\
 We can construct a generic, normalizable state by taking a superposition of these plane wave solutions obtained from the physical eigenstates, each of which obeys their corresponding Hamiltonian and momentum constraints,
\begin{equation}
\label{eq:KG_superposition}
\kett{\Psi} \: = \: \int d^{3}k \: c({\vec{k}}) \kett{\psi_{\vec{k}}} \: , 
\end{equation}
which written in the coordinate basis will be a correlated or entangled state between the temporal and spatial degrees of freedom\footnote{A similar entangled state $\kett{\Psi}$ can be written in the Page-Wootters formulation too with entanglement between states in $\hst$ and $\hs_S$ \cite{giovannetti2015quantum}.},
\begin{equation}
\label{eq:KG_gen_state}
\kett{\Psi} \: = \: \int d^{3}k \: \int \: dt \: d^{3}x \: c(\vec{k}) \exp{\left(-i E(\vec{k}) t + i \vec{k} \cdot \vec{x}\right)} \: \ket{t}\otimes \ket{\vec{x}} \: .
\end{equation}
Thus, time evolution and spatial translations can be interpreted in terms of entanglement between the spatial and temporal degrees of freedom in the global physical states $\kett{\Psi} \in \hs$. The corresponding wavefunction $\Psi(t,\vec{x})$ of the generic state $\kett{\Psi}$ can then be found by conditioning with a coordinate basis element $\ket{t}\otimes\ket{\vec{x}}$,
\begin{equation}
\Psi(t,\vec{x}) \equiv \bigg(\bra{t} \otimes \bra{\vec{x}}\bigg) \kett{\Psi} = \int d^{3}k  \: c(\vec{k}) \exp{\left(-i E(\vec{k}) t + i \vec{k} \cdot \vec{x}\right)}  \: ,
\end{equation}
which we recognize as the generic solution of the Klein-Gordon equation in the relativistic quantum mechanics of a particle. Issues regarding normalizability, particularly with respect to the temporal degree of freedom, and the use of continuous functional spaces and rigged Hilbert spaces in such a paradigm have been discussed elsewhere and the interested reader is encouraged to look at Refs.  \cite{giovannetti2015quantum, smith2019quantizing} (and references therein). We would however, like to point out that normalization in the temporal coordinate is a subtle issue and is treated rather distinctly than normalization over space. We feel this is an interesting point that warrants further investigation in future work to help understand it better. 
\\
\\
Thus, the physical Hilbert space $\hs_{\mrm{phys}}$ can be defined to be the span of physical eigenstates $\kett{\psi_{\vec{k}}}$ that satisfy the Hamiltonian and momentum constraint equations,
\begin{equation}
\label{eq:Hphys_KG}
\hs_{\mrm{phys}} \simeq \mrm{span}\left\{ \kett{\psi_{\vec{k}}} \: : \: \J_{H}(\vec{k})\kett{\psi_{\vec{k}}}  = \J_{P_{j}}(\vec{k})\kett{\psi_{\vec{k}}} = 0 \: \: \forall \: \vec{k} \in \mathbb{R}^{3}\: , \: j = x,y,z \right\} \: .
\end{equation}
We can also recover the Klein-Gordon equation explicitly by combining the Hamiltonian and momentum constraints of Eqs. (\ref{eq:KG_phys_states_H}) and (\ref{eq:KG_phys_states_mom}),
\begin{equation}
\label{eq:KG_eq_psi_k}
\left[\left( \pt \otimes \eye_{\vec{x}} \otimes \eye_{\mrm{spin}}\right)^{2} \: - \: \sum_{j = x,y,z}\left(\eye_{t} \otimes \hat{p}_{j} \otimes \eye_{\mrm{spin}}\right)^{2} \right] \kett{\psi_{\vec{k}}} \: = \: m^{2}\kett{\psi_{\vec{k}}} \: .
\end{equation}
Once we have the Klein-Gordon equation for a single physical eigenstate $\kett{\psi_{\vec{k}}}$ in the form of Eq. (\ref{eq:KG_eq_psi_k}), we can superpose appropriately and write a similar equation for an arbitrary physical state $\kett{\Psi}$ of Eq. (\ref{eq:KG_gen_state}),
\begin{equation}
\label{eq:KG_eq_noncov}
\left[\left( \pt \otimes \eye_{\vec{x}} \otimes \eye_{\mrm{spin}}\right)^{2} \: - \: \sum_{j = x,y,z}\left(\eye_{t} \otimes \hat{p}_{j} \otimes \eye_{\mrm{spin}} \right)^{2} \right] \kett{\Psi} \: = \: m^{2}\kett{\Psi} \: ,
\end{equation}
We can now use the relativistically covariant notation of Eq. (\ref{eq:P_mu}) and use the Minkowski flat metric $\eta^{\mu \nu} = \mrm{diag}\left(+1,-1,-1,-1\right)$ to give the Klein-Gordon equation a much more familiar form,
\begin{equation}
\bigg( \hat{P}^{\mu} \hat{P}_{\mu} \otimes \eye_{\mrm{spin}} \: - \: m^{2} \:\eye_{\hs}\bigg) \approx 0 \implies \bigg( \hat{P}^{\mu} \hat{P}_{\mu} \otimes \eye_{\mrm{spin}} \:-  \: m^{2} \: \eye_{\hs} \bigg) \kett{\Psi} \: = \: 0 \: , 
\end{equation}
where the repeated index $\mu$ is summed over. Thus, we see that we can recover the quantum mechanics of a spinless relativistic particle obeying the Klein-Gordon equation without having to deal with the ``square-root" Hamiltonian operator explicitly, but rather by working with a collection of \emph{linear, first-order} Hamiltonian and momentum constraints in a setup that deals with time and space on an equal footing in Hilbert space. It is important to note that relativistic dispersion relations were applied to eigenvalues featuring in the constraints and not be used as operator-values equations. While we obtain a collection of physical eigenstates $\kett{\psi_{\vec{k}}}$ for each spatial momentum $\vec{k}$ from the constraints, we are able to recover the full Klein-Gordon equation for generic physical states $\kett{\Psi} \in \hs$.
\\
\\
We now briefly discuss how multi-particle interactions can be accounted in such a relativistic, first-quantization, linear constraint formulation. Let us begin by describing two \emph{non-interacting} Klein-Gordon particles, and then introduce interactions. The Hilbert space for two such particles will have two copies of the spinorial Hilbert space, and therefore takes the form,
\begin{equation}
\hs \simeq \hst \otimes \hsx \otimes \hsspin \otimes \hsspin \: .
\end{equation}
Even though each $\hsspin$ in this Klein-Gordon case is one-dimensional, they are important for identification of the energy and momentum of each particle, as in the single particle case too. In the non-interacting case, both particles have a well-defined notion of their respective momenta $\vec{k}_{1}$ and $\vec{k}_{2}$ and energies $E_{1} = \sqrt{|\vec{k}_{1}|^{2} + m^{2}}$ and $E_{2} = \sqrt{|\vec{k}_{2}|^{2} + m^{2}}$. The total momentum of the system is then simply $\vec{k}_{\mathrm{tot}} = \vec{k}_{1} + \vec{k}_{2}$, and total energy $E_{\mathrm{tot}} = E_{1} + E_{2}$.	The Hamiltonian constraint of the system, for the total energy $E_{\mathrm{tot}}$ and momentum $\vec{k}_{\mathrm{tot}}$, can be written as,
\begin{equation}
\label{eq:KG_JH_2particle_noninter}
\hat{\mathbb{J}}_{H}(\vec{k}_\mathrm{tot}) = \pt \otimes \eye_{\vec{x}} \otimes \eye_{\mrm{spin}} \otimes  \eye_{\mrm{spin}}  \: + \: \eye_{t} \otimes \eye_{\vec{x}} \otimes \( E_{\mathrm{tot}}(\vec{k}_{1},\vec{k}_{2}) \: \eye_{\mrm{spin}} \otimes  \eye_{\mrm{spin}} \) \: , 
\end{equation}
and the momentum constraints,
\begin{equation}
\label{eq:KG_JPj_2particle_noninter}
\hat{\mathbb{J}}_{P_{j}}(\vec{k}_\mathrm{tot}) = \eye_{t} \otimes \hat{p}_{j} \otimes \eye_{\mrm{spin}} \: - \: \eye_{t} \otimes \eye_{\vec{x}} \otimes \(k_{\mathrm{tot},j} \: \eye_{\mrm{spin}} \otimes  \eye_{\mrm{spin}} \) \: \: \mrm{for} \:  j = x,y,z \: . 
\end{equation}
The physical states annihilated by these constraints have the form,
\begin{equation}
\kett{\psi_{\mathrm{tot}} (\vec{k}_{1},\vec{k}_{2})  }   = \ket{p_{t} = -E_{\mathrm{tot}}(\vec{k}_{1},\vec{k}_{2})} \otimes \ket{p_{x} = k_{\mathrm{tot},x}} \otimes \ket{p_{y} = k_{\mathrm{tot},y}} \otimes \ket{p_{z} = k_{\mathrm{tot},z}} \: ,
\end{equation}
which leads to an unentangled wavefunction of the two particles, as one would expect from two non-interacting particles,
\begin{equation}
\psi_{\mathrm{tot}} (\vec{k}_{1},\vec{k}_{2}) \: = \: \psi_{\vec{k}_{1}} \cdot  \psi_{\vec{k}_{2}} \: \sim \: \exp{\left(-i E_{1}(\vec{k}_{1}) t + i \vec{k}_{1} \cdot \vec{x}\right)}  \: \exp{\left(-i E_{2}(\vec{k}_{1}) t + i \vec{k}_{2} \cdot \vec{x}\right)} \: .
\end{equation}

Now, let us move on to the case of two \emph{interacting} particles. We will not write down particular forms of Hamiltonians which describe relativistically consistent interactions, but rather write down the constraints in terms of total energy and momentum. For a certain interacting system, each of the particles will typically \emph{not} have well defined notions of individual momenta and energies, but rather only total energy $E_{\mathrm{tot}}$ and total momentum $\vec{k}_{\mathrm{tot}}$ eigenvalues of the combined system which are its defining quantities (just like an operator is specified by its eigenvalue spectrum). The wavefunction of two such interacting particles will be an entangled state in the basis of their individual energy-momentum eigenstates,
\begin{equation}
\label{eq:KG_entangled}
\psi_{\mathrm{tot}} (\vec{k}_{1},\vec{k}_{2}) \: = \: \sum_{\vec{k}_{1},\vec{k}_{2}} c(\vec{k}_{1},\vec{k}_{2}) \exp{\left(-i E_{1}(\vec{k}_{1}) t + i \vec{k}_{1} \cdot \vec{x}\right)}  \: \exp{\left(-i E_{2}(\vec{k}_{1}) t + i \vec{k}_{2} \cdot \vec{x}\right)} \: ,
\end{equation}
where the basis states for each particle are labeled by $\vec{k}_{1}$ and $\vec{k}_{2}$ and corresponding energies $E_{1} = \sqrt{|\vec{k}_{1}|^{2} + m^{2}}$ and $E_{2} = \sqrt{|\vec{k}_{2}|^{2} + m^{2}}$. The total energy $E_{\mathrm{tot}}$ and total momentum $\vec{k}_{\mathrm{tot}}$ would typically not be be sums of individual particle contributions in the interacting case. The joint interacting wavefunction $\psi_{\mathrm{tot}} (\vec{k}_{1},\vec{k}_{2})$ of Eq. (\ref{eq:KG_entangled}) is annihilated by the same constraints of Eqs. (\ref{eq:KG_JH_2particle_noninter}) and (\ref{eq:KG_JPj_2particle_noninter}), and are eigenstates of total energy and total momentum u. The constraints, in this interacting case, give the translation equations,
\begin{equation}\label{eq:KG__inter_wavefunction_t}
i \frac{\partial}{\partial t} \psi_{\mathrm{tot}} (\vec{k}_{1},\vec{k}_{2}) \: = \:  E_{\mathrm{tot}} \: \psi_{\mathrm{tot}} (\vec{k}_{1},\vec{k}_{2}) \: .
\end{equation}
\begin{equation}\label{eq:KG_inter_wavefunction_x}
-i \vec{\nabla} \psi_{\mathrm{tot}} (\vec{k}_{1},\vec{k}_{2})\: = \:  \vec{k}_{\mathrm{tot}} \: \psi_{\mathrm{tot}} (\vec{k}_{1},\vec{k}_{2})\: ,
\end{equation}
which, when used with Eq. (\ref{eq:KG_entangled}) will yield a set of linear equations which can, in principle, be solved for $c(\vec{k}_{1},\vec{k}_{2})$ based on the total energy $E_{\mathrm{tot}}$ and total momentum $\vec{k}_{\mathrm{tot}}$. The essential idea here is that composite, interacting systems are defined by the collection of the spectrum of their total energy and momentum eigenvalues based on relativistically compatible interactions. In this linear constraint-based formulation, our aim is not to ``solve'' for these eigenvalues, but rather consistently write the constraint equations and base the total wavefunction as an entangled superposition of single particle, free basis functions using the total energy-momentum eigenvalues.
\\
\\
One can generalize this construction to include $N$-particles, by simply adding more copies of the spinorial Hilbert space, and appropriately generalizing the constraint and wavefunctions equations above,
\begin{equation}
\hs \simeq \hst \otimes \hsx \otimes \left(\hsspin \right)^{\otimes N}  \: .
\end{equation}
\\
\\
 While relativistically consistent interactions between multiple particles can be added in this scheme as described, interaction of a single particle with a background field are typically not relativistically compatible since they break either time or space translational symmetry that leads to the energy and/or momentum of the particle not being conserved. Hence, we do not add interaction with a background field in the constraint equations above since they would not be consistent with the commuting compatibility $\lcb \J_{H}, \J_{P_{j}} \rcb = 0$ of the constraints. It is therefore instructive to investigate the non-relativistic limit of the Klein-Gordon setup and see how one can restore interactions to recover the non-relativistic Schr\"odinger equation in the conventional form. For Eq. (\ref{eq:KG_phys_states_H}), we can take the non-relativistic limit ($|\vec{k}| << m$) by expanding $E(\vec{k})$ in powers of $|\vec{k}|^{2}$ and retaining the leading order $\vec{k}$-dependent piece along with dropping the constant rest mass energy $m$ contribution,
\begin{equation}
\left(\pt \otimes \eye_{\vec{x}} \otimes 1 \: + \: \eye_{t} \otimes \eye_{\vec{x}} \otimes \frac{|\vec{k}|^{2}}{2m} \right) \kett{\psi_{\vec{k}}}_{\mrm{NR}} = 0 \: , 
\end{equation}
where $\kett{\psi_{\vec{k}}}_{\mrm{NR}}$ is the non-relativistic physical eigenstate. Combining with the momentum constraint of Eq. (\ref{eq:KG_phys_states_mom}), this yields the Schr\"odinger equation for a given non-relativistic physical eigenstate in the full Hilbert space $\hs$,
\begin{equation}
\(\pt \otimes \eye_{\vec{x}} \otimes 1\) \kett{\psi_{\vec{k}}}_{\mrm{NR}} \: = \: -\[\sum_{j = x,y,z} \frac{\left(\eye_{t} \otimes \hat{p}_{j} \right)^{2}}{2m}\]\kett{\psi_{\vec{k}}}_{\mrm{NR}} \: .
\end{equation}
We can now construct a general state by superposition of different physical eigenstates as in Eq. (\ref{eq:KG_superposition}) that gives us the Schr\"odinger equation for a non-relativistic free particle. At this stage, since we are no longer working to keep our constraints relativistically compatible, we can also add by-hand an ``interaction term" $V(t,\vec{x})$ to model interactions of the non-relativistic particle with a background field, 
\begin{equation}
\(\pt \otimes \eye_{\vec{x}} \otimes 1\) \kett{\Psi}_{\mrm{NR}} \: = \: -\[\sum_{j = x,y,z} \frac{\left(\eye_{t} \otimes \hat{p}_{j} \right)^{2}}{2m}\ \: + \: V\(\toptr\otimes \eye_{\vec{x}} \:, \: \eye_{t}\otimes\hat{\vec{x}}\) \]\kett{\Psi}_{\mrm{NR}} \: ,
\end{equation}
which when written in terms of the wavefunction, gives us,
\begin{equation}
\label{eq:KG_NR_wavefunction_k}
i \frac{\partial}{\partial t} \Psi_{\mrm{NR}}(t,\vec{x}) \: = \:\( -\frac{1}{2m}{\vec{\nabla}}^{2} + V(t,\vec{x})\) \: \Psi_{\mrm{NR}}(t,\vec{x}) \: .
\end{equation} 
We would again like to emphasize here that interactions could only be added in the non-relativistic limit where the Hamiltonian and momentum need not commute. We briefly remark on this aspect about interactions and its connection with quantum field theory in section \ref{sec:discussion}.
\subsection{Dirac Equation (Spin-1/2)} \label{sec:Dirac}
Now that we have showed a constraint-based approach to Klein-Gordon equation of a single relativistic particle of spin-0, let us focus on the Dirac equation which describes fermionic particles with spin-1/2. Since one can find an instructive treatment of the Dirac equation in most advanced undergraduate quantum mechanics texts, we will not reproduce that discussion here. Instead, we will jump right in to describe using Hamiltonian and momentum constraints to work out the Dirac equation. One key thing to remember is that even in the usual textbook construction, the Dirac equation -- at variance with the standard treatment of the Klein-Gordon equation -- is a linear, first-order equation in both energy and momentum (or one may say, first-order in its treatment of time and space).
\\
\indent For relativistic quantum mechanics of a spin-1/2 particle, we know that the dimension of the spinorial Hilbert space $\hsspin$ is $\dim\hsspin = 4$ since it is used to describe both the particle and its antiparticle (which could be the same as the particle itself, as in the case of Majorana fermions). Following the usual construction, we use gamma matrices $\gamma^{\mu}$ of spinors which satisfy the anti-commutation relations of Clifford algebra, $\{ \gamma^{\mu} , \gamma^{\nu} \} = 2 \eta^{\mu \nu} \: \eye_{\mrm{spin}}$,  where $\{A,B\}  = AB+ BA$ is the anti-commutator of $A$ and $B$, and we have used the metric signature $\eta^{\mu \nu} = \mrm{diag}\left(+1,-1,-1,-1\right)$. Depending on the representation of the gamma matrices, we can describe both Dirac fermions (particles having distinct antiparticles) and Majorana fermions (particles which are their own antiparticles) \cite{Pal:2010ih}. Let us define,
\begin{equation}
\dmat{\beta} \equiv \gamma^{0} \: \: \:, \: \: \: \dmat{\alpha_{j}} \equiv \gamma^{0}\gamma^{j} \:, \: \:   j = x,y,z \: , 
\end{equation}
which satisfy $\alpha^{2}_{j} = \eye_{\mrm{spin}} \: \forall j$, $\beta^{2} = \eye_{\mrm{spin}}$, $\{ \alpha_{j}, \alpha_{{j}'} \} = 0  \: \forall j \neq {j}'$ and $\{\alpha_{j}, \beta\} = 0 \: \forall \: j$.
As before, the Hamiltonian and momentum constraints will be parametrized by a vector $\vec{k} \in \mathbb{R}^{3}$. The Hamiltonian constraint is given by,
\begin{equation}
 \label{eq:Dirac_JH}
\hat{\mathbb{J}}_{H}(\vec{k}) = \pt \otimes \eye_{\vec{x}} \otimes \eye_{\mrm{spin}} \: + \: \eye_{t} \otimes \eye_{\vec{x}} \otimes \(\sum_{j=x,y,z} k_{j} \dmat{\alpha_j} + m \dmat{\beta}\) \: , 
\end{equation}
and the momentum constraint by,
\begin{equation}
\label{eq:Dirac_JPj}
\hat{\mathbb{J}}_{P_{j}}(\vec{k}) = \eye_{t} \otimes \hat{p}_{j} \otimes \eye_{\mrm{spin}} \: - \: \eye_{t} \otimes \eye_{\vec{x}} \otimes \(k_{j}\: \eye_{\mrm{spin}}\) \: \: \mrm{for} \:  j = x,y,z \: . 
\end{equation}
Notice, in the constraint equations above, the values of energy and momentum eigenvalues are associated explicitly with operators that act on $\hsspin$, just how we did in the Klein-Gordon case (though is was a trivial association there). The spinorial term in the Hamiltonian constraint can be identified as the matrix square root of the Klein-Gordon energy-momentum dispersion relation,
\begin{equation}
\(\sum_{j=x,y,z} k_{j} \dmat{\alpha_j} + m \dmat{\beta}\)^{2} \: = \: \(|\vec{k}|^{2} + m^{2}\) \: \eye_{\mrm{spin}} \: .
\end{equation}
This is expected, since the Dirac equation, even in the standard construction, is understood as a ``square-root" of the Klein-Gordon equation and doing so necessarily relies on the spin of the particle.
  Physical eigenstates $\kett{\psi_{\vec{k}}}$ are identified to be the ones that are annihilated by the constraints,
  \begin{equation}
\label{eq:Dirac_phys_states_H}
\J_{H}(\vec{k}) \approx 0 \: \implies \:  \J_{H}(\vec{k}) \: \kett{\psi_{\vec{k}}} = 0 \: , 
\end{equation}
\begin{equation}
\label{eq:Dirac_phys_states_mom}
\J_{P_{j}}(\vec{k}) \approx 0 \: \implies \: \J_{P_{j}}(\vec{k}) \: \kett{\psi_{\vec{k}}} = 0 \: \: \mrm{for} \:  j = x,y,z \: . 
\end{equation}
One can now combine the Hamiltonian and momentum constraints of Eqs. (\ref{eq:Dirac_phys_states_H}) and (\ref{eq:Dirac_phys_states_mom}) to eliminate the explicit $\vec{k}$ dependence,
\begin{equation}
\label{eq:Dirac_eq_1}
\pt \otimes \eye_{\vec{x}} \otimes \eye_{\mrm{spin}} \: + \: \sum_{j=x,y,z}\bigg( \eye_{t} \otimes \hat{p}_{j} \otimes \dmat{\alpha_j} \bigg) \: + \: \(\eye_{t} \otimes \eye_{\vec{x}} \otimes m\: \dmat{\beta}\) \: \approx \: 0 \: ,
\end{equation}
which is precisely the Schr\"odinger equation for a spin-1/2 particle, which in the conventional form is written as,
\begin{equation}
i\frac{\partial}{\partial t} \psi \: = \: \bigg( {\dmat{\vec{\alpha}}} \cdot \hat{\vec{p}}  + m\dmat{\beta}\bigg) \psi \: . 
\end{equation}
We can also recover the Dirac equation by pre-multiplying Eq. (\ref{eq:Dirac_eq_1}) with $\bigg(\eye_{t} \otimes \eye_{\vec{x}} \otimes \: \dmat{\beta}\bigg) $,
\begin{equation}
\label{eq:Dirac_eq_2}
\(\pt \otimes \eye_{\vec{x}} \otimes \dmat{\beta}\) \: + \: \sum_{j=x,y,z}\bigg( \eye_{t} \otimes \hat{p}_{j} \otimes \dmat{\beta} \: \dmat{\alpha_j} \bigg) \: + \: m \: \bigg(\eye_{t} \otimes  \eye_{\vec{x}} \otimes \eye_{\mrm{spin}}\bigg) \: \approx \: 0 \: .
\end{equation}
Switching back to the gamma matrix notation,
\begin{equation}
\dmat{\gamma^{0}} \equiv \dmat{\beta} \: \: \:, \: \: \: \dmat{\gamma^{j}} \equiv \dmat{\beta} \dmat{\alpha_{j}} \:, \: \:   j = x,y,z \: , 
\end{equation}
and this gives us the Dirac equation,
\begin{equation}
\(\pt \otimes \eye_{\vec{x}} \otimes \dmat{\gamma^{0}}\) \: + \: \sum_{j=x,y,z}\bigg( \eye_{t} \otimes \hat{p}_{j} \otimes \dmat{\gamma^{j}} \bigg) \: + \: m \: \eye_{\hs} \: \approx \: 0 \: .
\end{equation}
We can now write this equation in the relativistically covariant notation of Eq. (\ref{eq:P_mu}),
\begin{equation}
\label{eq:dirac_eq}
\hat{P}_{\mu} \otimes \dmat{\gamma^{\mu}} \: + \: m \: \eye_{\hs} \: \approx \: 0 \: .
\end{equation}
where the repeated index $\mu$ is summed over. One can introduce interactions with a background electromagnetic field $A_{\mu}(t,\vec{x})$ by adding it as an effective term at the level of the Dirac equation of Eq. (\ref{eq:dirac_eq}). Adding such an interaction with a background field directly in the constraint equations of Eqs. (\ref{eq:Dirac_JH}) and (\ref{eq:Dirac_JPj}) will be inconsistent since it will break spatial/temporal translation symmetry for the relativistic particle whose physics we are focusing on. One can envision assigning Hilbert spaces to the background field as quantum degrees of freedom and then adding relativistic compatible interactions with other fields. We will briefly discuss this point from a field-theoretic point of view in section \ref{sec:discussion}.
\\
\\
\indent Thus, we see that we are able to treat both Klein-Gordon and Dirac equations with a common approach based on linear, first-order constraints by treating space and time on an equal footing in Hilbert space and applying dispersion relations to eigenvalues instead of an operator-valued equation in $\hsx$.
\section{Symmetry transformations as Basis Changes}\label{sec:basis_changes}
Treating space and time on an equal footing as quantum degrees of freedom in Hilbert space can help us analyse symmetry transformations using unitary basis changes in Hilbert space. This gives a stronger quantum mechanical ground for symmetry transformations, especially in relativistic quantum mechanics where transformations of temporal degrees of freedom are often handled in an ad-hoc, often classical way compared to the spatial degrees of freedom. It also lets us tie together global and local symmetry transformations into one framework and in this section, we will focus on two important symmetry transformations: Lorentz transformations and $U(1)$ gauge symmetry. Global symmetries, such as Lorentz transformations, will affect decomposition changes in Hilbert space, whereas local symmetries, such as $U(1)$, will correspond to basis changes in Hilbert space while leaving the decomposition invariant.
\subsection{Lorentz Transformations}
Lorentz transformations are global transformations that mix space and time components of 4-vectors while preserving the speed of light, the causal structure and the form of laws of physics in each inertial reference frame. In Hilbert space, we will implement a Lorentz transformation as a decomposition change of $\hst\otimes \hsx$. Thus, it is an overall basis change that alters the factorization between the temporal and spatial degrees of freedom in Hilbert space. We will implement a Lorentz transformation $\Lambda$ by a unitary transformation $\Ult(\Lambda)$ that changes the decomposition,
\begin{equation}
\Ult(\Lambda) \: : \: \hst \otimes \hsx \: \rightarrow \:  {\hst}' \otimes {\hsx}' \: .
\end{equation}
Such a decomposition change manifests itself by mixing conjugate operators in $\hst$ and $\hsx$ under the following transformation,
\begin{equation}
\label{eq:Xmu_LT}
{\hat{X}^{\mu}}' \: = \Ult^{\dag}(\Lambda) \:  \hat{X}^{\mu} \:  \Ult(\Lambda) \: = \: \Lambda^{\mu}_{\nu} \: \hat{X}^{\nu} \: ,
\end{equation}
\begin{equation}
{\hat{P}}'_{\mu} \: = \Ult^{\dag}(\Lambda) \:  \hat{P}_{\mu} \: \Ult(\Lambda) \: = \: \Lambda_{\mu}^{\nu} \: \hat{P}_{\nu} \: .
\end{equation}
It is important to note that ${\hat{X}^{\mu}}'$ and $\hat{P}'_{\mu}$ are ``separable" operators \textit{i.e.} they have a tensor product structure in the transformed space ${\hst}' \otimes {\hsx}'$. For example, the components of ${\hat{X}^{\mu}}' \in \mathcal{L}({\hst}' \otimes {\hsx}')$ written explicitly are, 
\begin{align}
{\hat{X}^{0}}' \equiv \toptr' \otimes \eye'_{\vec{x}}, \: \: \: \: \: \: {\hat{X}^{1}}' \equiv {\eye_{t}} \otimes \x' \otimes {\eye_{y}}' \otimes {\eye_{z}}' 
,\\
 {\hat{X}^{2}}' \equiv \eye'_{t} \otimes {\eye_{x}}' \otimes \y' \otimes {\eye_{z}}' 
, \: \: \: \: \: \: {\hat{X}^{3}}' \equiv \eye'_{t} \otimes {\eye_{x}}' \otimes {\eye_{y}}' \otimes {\z}' \: ,
\end{align} 
and therefore, each of these transformed operators is a linear combination of separable operators in $\hst\otimes \hsx$ as governed by Eq. (\ref{eq:Xmu_LT}) (and hence, ${\hat{X}^{\mu}}'$ and ${\hat{P}}'_{\mu}$ are themselves are not separable in $\hst\otimes \hsx$).
\\
\\
Let us look at this in more detail with an example of the Klein-Gordon equation as treated in section \ref{sec:KG} to study how the Hamiltonian and momentum constraints transform under a Lorentz transformation and its implications. For concreteness and simplicity, let us focus on a Lorentz boost with a boost parameter $v$ (the relative speed of the two inertial frames in units with $c = 1$) along the $x$-direction. We define $\Gamma = \(1 - v^{2}\)^{-1/2}$, and with this, the Lorentz transformation matrix $\Lambda$ takes the form,
\begin{equation}
\Lambda^{\mu}_{\nu} \: = \:
\begin{bmatrix}
 \Gamma & -\Gamma v & 0 \:  & 0 \\
-\Gamma v & \Gamma  & 0 \:  & 0 \\
0 & 0 & 0 \:  & 0\\
0 & 0 & 0 \:  & 0\\
\end{bmatrix} \: .
\end{equation}
In this case, the unitary transform $\Ult(\Lambda)$ changes the decomposition of $\hst \otimes \hs_{x} $ to ${\hst}' \otimes {\hs_{x}}'$ and leaves $\hs_{y} \otimes \hs_{z}$ untransformed, such that,

\begin{equation}
\eye_{t} \otimes \eye_{x} \otimes \py \otimes \eye_{z} \: \: \longmapsto \: \: {\eye_{t}}' \otimes {\eye_{x}}' \otimes {\py}' \otimes {\eye_{z}}' \: \: = \: \: \eye_{t} \otimes \eye_{x} \otimes \py \otimes \eye_{z} \:,
\end{equation}
\begin{equation}
\eye_{t} \otimes \eye_{x} \otimes \eye_{y} \otimes \pz \: \: \longmapsto \: \: {\eye_{t}}' \otimes {\eye_{x}}' \otimes {\eye_{y}} \otimes {\pz}' \: \: = \: \: \eye_{t} \otimes \eye_{x} \otimes \eye_{y} \otimes \pz \: ,
\end{equation}
\begin{equation}
\pt \otimes \eye_{\vec{x}} \: \: \longmapsto  \: \: {\pt}' \otimes {\eye_{\vec{x}}}' \: \: = \: \:  \Gamma \({\pt}' \otimes {\eye_{\vec{x}}}'\) \: - \: \Gamma v \({\eye_{t}}' \otimes {\hat{p}_{x}}' \otimes {\eye_{y}} \otimes {\eye_{z}}\)
\end{equation}
\begin{equation}
\eye_{t} \otimes \px \otimes \eye_{y}\otimes \eye_{z} \: \: \longmapsto  \: \: {\eye_{t}}' \otimes {\px}' \otimes \eye_{y}\otimes \eye_{z} \: \: = \: \:   \Gamma \({\eye_{t}}' \otimes {\px}' \otimes \eye_{y} \otimes \eye_{z} \) \: - \: \Gamma v \( {\pt}' \otimes {\eye_{x}}' \otimes {\eye_{y}} \otimes {\eye_{z}}\) \: .
\end{equation}
We can now see from the above equations that a Lorentz transformation changes the decomposition of Hilbert space. Under this transformation, the Hamiltonian constraint of Eq. (\ref{eq:KG_JH}) transforms as,
\begin{equation}
\label{eq:KG_JH_LT}
{\J_{H}}'(\vec{k}) \: = \: \Gamma \({\pt}' \otimes {\eye_{\vec{x}}}' \otimes 1\) \: - \: \Gamma v \({\eye_{t}}' \otimes {\hat{p}_{x}}' \otimes {\eye_{y}} \otimes {\eye_{z}} \otimes 1 \) \: + \: \({\eye_{t}}'\otimes {\eye_{\vec{x}}}'\otimes E(\vec{k}) \) \: \approx \: 0 \: , 
\end{equation}
and the $x$-momentum constraint takes the form,
\begin{equation}
\label{eq:KG_Jpx_LT}
{\J_{P_{x}}}'(\vec{k}) \: = \: \Gamma \({\eye_{t}}' \otimes {\px}' \otimes \eye_{y} \otimes \eye_{z} \otimes 1\) \: - \: \Gamma v \( {\pt}' \otimes {\eye_{x}}' \otimes {\eye_{y}} \otimes {\eye_{z}} \otimes 1\) \: - \: \( {\eye_{t}}' \otimes {\eye_{\vec{x}}}' \otimes k_{x} \)\: \approx \: 0 \: .
\end{equation}
The $y$ and $z$-momentum constraints remain unchanged since the Lorentz transformation only mixes $\hst\otimes\hs_{x}$. As expected, the Hamiltonian and momentum constraints of Eqs. (\ref{eq:KG_JH_LT}) and (\ref{eq:KG_Jpx_LT}) have mixed terms but we can decouple them by substitution of one equation into the other. The transformed Hamiltonian constraint then becomes,
\begin{equation}
{\J_{H}}'(\vec{k}) \: = \: \({\pt}' \otimes {\eye_{\vec{x}}}' \otimes 1\) \: + \: \({\eye_{t}}' \otimes {\eye_{\vec{x}}}' \otimes \(\Gamma E(\vec{k}) - \Gamma v \: k_{x} \)\) \:\approx \: 0 \: , 
\end{equation}
which, as expected, represents a Hamiltonian constraint with the Lorentz transformed energy ${E}' = \(\Gamma E(\vec{k}) - \Gamma v \: k_{x} \)$. The momentum constraint similarly transforms to reflect the Lorentz transformed momentum ${}k_{x}' = \(\Gamma k_{x} - \Gamma v \: E(k)\)$,
\begin{equation}
{\J_{P_{x}}}'(\vec{k}) \: = \: \({\eye_{t}}' \otimes {\px}' \otimes \eye_{y} \otimes \eye_{z} \otimes 1\) \: - \: \({\eye_{t}}' \otimes {\eye_{\vec{x}}}' \otimes \(\Gamma k_{x} - \Gamma v \: E(k)\)\) \:\approx \: 0 \: 
\end{equation}
\\
Thus, we see that Lorentz transformations that are global symmetry transformations are implemented as global basis changes in Hilbert space altering the factorization between the temporal and spatial factors of Hilbert space. One can perform a similar Lorentz transformation for the Dirac equation as discussed in section \ref{sec:Dirac} though we avoid repeating a similar analysis here.
\subsection{U(1) Symmetry}
Let us now look at $U(1)$ gauge symmetry though the lens of unitary basis changes in Hilbert space. The $U(1)$ gauge symmetry is a local transformation that lets the wavefunction pick up a local phase, 
\begin{equation}
\label{eq:U1_basic}
\Psi(t,\vec{x}) \: \longmapsto \: {\Psi}'(t,\vec{x}) \: = \:  e^{i \lambda(t,\vec{x})} \psi(t,\vec{x}) \: ,
\end{equation}
for a gauge function\footnote{The function $\lambda(t,\vec{x})$ is typically taken to be continuous and sufficiently differentiable in its variables and dies off rapidly enough as $\vec{x} \rightarrow \pm \infty$.} $\lambda(t,\vec{x})$. The usual story of the $U(1)$ transformation is told in the presence of a gauge field $A_{\mu}(X)$ to which the particle couples. Under the gauge transformation of Eq. (\ref{eq:U1_basic}), we require the gauge field $A_{\mu}$ to transform accordingly,
\begin{equation}
\label{eq:U1_Avec}
\vec{A}(t,\vec{x}) \: \longrightarrow \: \vec{A}(t,\vec{x}) - \vec{\nabla}\lambda(t,\vec{x}) \:,
\end{equation}
\begin{equation}
\label{eq:U1_A0}
A_{0}(t,\vec{x}) \: \longrightarrow \: A_{0}(t,\vec{x}) + \frac{\partial}{\partial t} \lambda(t,\vec{x}) \: ,
\end{equation}
to keep the Schr\"odinger equation invariant. 
One of the outcomes of this transformation $A_{\mu}$ is to effectively shift the conjugate momentum operator,
\begin{equation}\label{eq:iddx}
\hat{\vec{p}} \: \longrightarrow \: \hat{\vec{p}} \: - \: \vec{\nabla}\lambda(t,\vec{x}) \: \implies \: -i\vec{\nabla} \longrightarrow -i\vec{\nabla} \: - \: \vec{\nabla}\lambda(t,\vec{x}) \: .
\end{equation}
On the other hand, due to lack of a conjugate temporal momentum in the textbook construction, the time derivative operator $i\partial_{t}$ (which equates itself to the Hamiltonian in the Schr\"odinger equation for physical states) therefore transforms as, 
\begin{equation}\label{eq:iddt}
i\frac{\partial}{\partial t} \longrightarrow i\frac{\partial}{\partial t} \: + \: \frac{\partial}{\partial t} \lambda(t,\vec{x}) \: .
\end{equation}
We now show, that in our construction which treats space and time on an equal footing in Hilbert space using linear, first-order constraints, $U(1)$ gauge transform is a local unitary transformation in the spatio-temporal Hilbert space $\hst\otimes \hsx$. Transformations, both in the temporal and spatial components, emerge naturally as a consequence of this unitary transformation. While one can couple the particle to an external/background field $A_{\mu}$, reference to this gauge field (which defines the field strength that is invariant under $U(1)$) is not explicitly required to affect the symmetry transformation. Spatial and temporal quantum degrees of freedom transform under a common mechanism, unlike as done in Eqs. (\ref{eq:iddx}) and (\ref{eq:iddt}). 
\\
\indent The unitary transformation $\hat{U}_{1} \in \mathcal{L}(\hst \otimes \hsx)$ that affects this $U(1)$ symmetry is not a global decomposition change in $\hst\otimes \hsx$, but rather a local basis change as one would expect from a gauge transformation and hence does not alter the decomposition of Hilbert space. The local nature of the unitary basis change operator is reflected it being diagonal in the coordinate basis $\{\ket{t}\otimes \ket{\vec{x}} \equiv \kett{t,\vec{x}} \}$ (we are considering a spinless particle for this analysis here),
\begin{equation}
\hat{U}_{1} \: = \: \( \int dt \: d^{3}x \: e^{-i \lambda(t,\vec{x})} \: \kett{t,\vec{x}} \braa{t,\vec{x}}\) \otimes \eye_{\mrm{spin}} \: = \: \exp{\( - i \lambda(\toptr\otimes \eye_{\vec{x}} , \eye_{t}\otimes \hat{\vec{x}}) \)} \otimes \eye_{\mrm{spin}} \: .
\end{equation}
Under this unitary transformation, the a state $\kett{\Psi} \in \hst \otimes \hsx$ transforms as follows,
\begin{equation}
\kett{{\Psi}'} \: = \: \hat{U}^{\dag}_{1} \kett{\Psi} \: ,
\end{equation}
which transforms the wavefunction as required for a $U(1)$ transformation by picking up a local phase,
\begin{equation}
{\Psi}'(t,\vec{x}) \equiv \bigg(\bra{t} \otimes \bra{\vec{x}}\bigg) \kett{{\Psi}'} \: = \: \bigg(\bra{t} \otimes \bra{\vec{x}}\bigg)\hat{U}^{\dag}_{1} \kett{\Psi} \: = \: e^{i\lambda(t,\vec{x})} \Psi(t,\vec{x}) \: .
\end{equation} 
The $\hat{U}_1$ transformation also transforms operators in a way consistent with a local gauge transformation. In this case, when time and space are treated on an equal footing in Hilbert space, we don't need to explicitly rely on the existence of a gauge field $A_{\mu}$ since the unitary transformation directly leads to transformation of the conjugate momenta in both $\hst$ and $\hsx$. Whereas in the textbook story, time is treated as an external parameter and not as a quantum degree of freedom, and hence there is no momenta conjugate to a time coordinate. Because of this, we have to impose transformations on the gauge field $A_{\mu}$ of Eqs. (\ref{eq:U1_Avec}) and (\ref{eq:U1_A0}) to keep evolution equations invariant. Here, we treat space and time on an equal footing and it is reflected in the unitary transformation of the conjugate momenta as follows,
\begin{equation}
{\hat{P}_{\mu}}' \: = \: \hat{U}^{\dag}_{1} \: \hat{P}_{\mu} \: \hat{U}_{1} \: = \: 
\exp{\( i \lambda(\hat{X}) \)} \: \hat{P}_{\mu} \: \exp{\(- i \lambda(\hat{X}) \)} \: ,
\end{equation} 
where $\lambda(\hat{X})$ is to denote that the function $\lambda$ depends on the coordinate operators $\hat{X}^{\mu}$ of Eq. (\ref{eq:X_mu}). We can use Baker-Campbell-Hausdorff lemma to further write,
\begin{equation}
\label{eq:BCH_U1}
{\hat{P}_{\mu}}' \: = \: \hat{P}_{\mu} \: + \: i\lcb \lambda(\hat{X}) , \hat{P}_{\mu} \rcb \: - \:  \frac{1}{2} \lcb \lambda(\hat{X}),\lcb \lambda(\hat{X}) , \hat{P}_{\mu} \rcb\rcb \: + \: \ldots \: . 
\end{equation}
Recalling that $\lcb \lambda(\hat{X}), \hat{P}_{\mu} \rcb \: = \: i \: \partial_{\mu}\lambda(\hat{X})$, we can further simplify the above equation to yield,
\begin{equation}
{\hat{P}_{\mu}}' \: = \: \hat{P}_{\mu} \: - \: \partial_{\mu}\lambda(\hat{X}) \: ,
\end{equation}
since two-point and higher nested commutators in Eq. (\ref{eq:BCH_U1}) all vanish because $\lambda(\hat{X})$ and its derivatives $\partial_{\mu}\lambda(\hat{X})$ are only functions of $\hat{X}_{\mu}$.  Thus, both spatial and temporal conjugate momenta get modified, in particular,
\begin{equation}
\label{eq:pt_U1}
{\pt}'\otimes \eye_{\vec{x}} \: = \: \pt \otimes \eye_{\vec{x}} \: - \: \frac{\partial}{\partial t} \lambda\(\toptr\otimes \eye_{\vec{x}} , \eye_{t}\otimes \hat{\vec{x}}\) 
\: ,
\end{equation}
and,
\begin{equation}
\label{eq:pvec_U1}
\eye_{t} \otimes {\hat{\vec{p}}}\:' \: = \: \eye_{t} \otimes {\hat{\vec{p}}} \: - \: \vec{\nabla}\lambda\(\toptr\otimes \eye_{\vec{x}} , \eye_{t}\otimes \hat{\vec{x}}\) 
\: .
\end{equation}
The coordinate operators $\hat{X}^{\mu}$ themselves do not transform since they commute with the local function $\lambda(\hat{X})$ and it should also be pointed out that by virtue of $\hat{U}_{1}$ being unitary, the CCR between the conjugate operators is left unmodified under the transformation. \\
\\
We can see how the Hamiltonian and momentum constraints transform under the $U(1)$ gauge transformation. Using Eqs. (\ref{eq:KG_JH}) and (\ref{eq:pt_U1}), the transformed Hamiltonian constraint operator looks like,
\begin{equation}
{\hat{\mathbb{J}}_{H}}'(\vec{k}) = \pt \otimes \eye_{\vec{x}} \otimes \eye_{\mrm{spin}} \: + \: \(\eye_{t} \otimes \eye_{\vec{x}} \otimes \(E(\vec{k}) \: \eye_{\mrm{spin}} \)  - \: \frac{\partial}{\partial t} \lambda\(\toptr\otimes \eye_{\vec{x}} , \eye_{t}\otimes \hat{\vec{x}}\)  \) \:   , 
\end{equation}  
which is the equivalent of the transformation of Eq. (\ref{eq:iddt}) but now arrived at by directly transforming the temporal conjugate momentum.
The transformed momentum constraint operator, using Eqs. (\ref{eq:KG_JPj}) and (\ref{eq:pt_U1}) becomes,
\begin{equation}
{\hat{\mathbb{J}}_{P_{j}}}'(\vec{k}) = \eye_{t} \otimes \hat{p}_{j} \otimes \eye_{\mrm{spin}} \: - \: \( \eye_{t} \otimes \eye_{\vec{x}} \otimes \( k_{j} \: \eye_{\mrm{spin}} \)   \: + \: \vec{\nabla}\lambda\(\toptr\otimes \eye_{\vec{x}} , \eye_{t}\otimes \hat{\vec{x}}\)   \)\:  \approx \: 0 \: \: \: \mrm{for} \:  j = x,y,z \: ,
\end{equation}
which is the equivalent of the transformation of Eq. (\ref{eq:iddx}) but now arrived at by directly transforming the spatial conjugate momentum by a unitary transformation on an equal footing with its temporal component. Thus, the evolution equations (both spatial and temporal) are transformed in accordance with a $U(1)$ unitary transformation on an equal footing. The sign difference between the space and time components of Eqs. (\ref{eq:U1_Avec}) and (\ref{eq:U1_A0}) are therefore traced back to the difference in the corresponding sign between the Hamiltonian and momentum constraints and not in the unitary transformation of the conjugate momenta. The transformed constraint operators still commute and the constraint equations are satisfied for the transformed state $\kett{{\Psi}'}$ by the transformed operators, \textit{i.e.} ${\hat{\mathbb{J}}_{H}}'(\vec{k})\kett{{\Psi}'} \:  = \: 0 \: = \: {\hat{\mathbb{J}}_{P_{j}}}'(\vec{k}) \kett{{\Psi}'}$, which is the statement that the evolution equations are left invariant under the $U(1)$ gauge transformation as expected.
\section{Discussion}\label{sec:discussion}
The quantum nature of space and time forms a core question in our understanding of quantum gravity. Motivated by considerations in a quantum-first approach to quantum gravity, we attempted to treat space and time on an equal footing in Hilbert space and focus on a paradigm dealing with linear, first-order constraints. Using both Hamiltonian and momentum constraints that annihilate physical states in Hilbert space, we can get emergent features like spatial and temporal translations.  Using these constraints, we analysed Klein-Gordon and Dirac equations and showed our analysis treats both equations with a uniform approach, arguing that dispersion relations should apply to eigenvalues and not be used as operator-valued equations. With such an approach, the ``square root" Hamiltonian in the Klein-Gordon theory is handled naturally on a common footing with the Dirac equation. Treating both time and space as quantum degrees of freedom in Hilbert space, the quantum mechanical status of Lorentz transformations and $U(1)$ symmetry is seen as change of basis or decomposition of Hilbert space. Our construction in this paper keeps space, time, and spin on an equal footing in Hilbert space as given by Eq. (\ref{eq:HS_decomposition}) and this gives a homogeneous use of the same underlying algebra operating in each of the Hilbert space factors. The Generalized Clifford Algebra \cite{Jagannathan:2010sb,Singh:2018qzk,SanthanamTekumalla1976} can be seen to provide this common algebraic structure for conjugate operators in $\hst$, $\hsx$ and $\hsspin$. In both $\hst$ and $\hsx$, it provides the conjugate algebra through Weyl's form of the CCR \cite{weyl1950theory} that approaches the Heisenberg CCR in the infinite-dimensional limit, and for $\hsspin$, it furnishes the spinor matrices obeying Clifford algebra (for instance, the Pauli matrices are the algebra obtained by a Generalized Clifford Algebra with two generators in two-dimensions). In addition, due to the Bekenstein bound \cite{Bekenstein:1980jp} and holographic principle \cite{Hooft:1993gx,Susskind:1994vu}, the Hilbert space of quantum gravity may be locally finite-dimensional \cite{Bao:2017rnv,Banks2000,Fischler2000}. In such a finite-dimensional scenario, the Generalized Clifford Algebra also offers a finite-dimensional version of conjugate operators obeying Weyl's exponential form of the CCR.
\\
\\
Such a program can be extended into various future directions, some of which we'd like to point out here and discuss their implications. One of the most natural and interesting generalizations of this approach is to formulate quantum field theory in this language. To this extent, we can imagine a Hilbert space decomposition, as an extension of Eq. (\ref{eq:HS_decomposition}),
\begin{equation}
\label{eq:QFT_HS}
\hs \: \simeq \: \hst \otimes \hsx \otimes \hs_{\mrm{matter}} \: ,
\end{equation}
which could describe quantum-mechanical matter $\hs_{\mrm{matter}}$ living on a background spacetime described by quantum degrees of freedom $\hst \otimes \hsx$. This represents a second quantized system, and is distinct from the first quantized systems discussed in the bulk of this paper. The structure of the spacetime Hilbert space in the context of field theory would warrant further investigation and its interplay with the matter Hilbert space. In the conventional field theory paradigm, one colloquially associates a Hilbert space at each point in space, and states in this space evolve unitarily through time. While the analysis in this paper focussed on Hamiltonian and momentum constraints parametrized by $\vec{k} \in \mathbb{R}^{3}$, as in Eqs. (\ref{eq:KG_phys_states_H}) and (\ref{eq:KG_phys_states_mom}), an extension to a field theory-like setup would allow writing single Hamiltonian and momentum constraints using the generators of the Poincar\'e group,
\begin{equation}
\label{eq:QFT_JH}
\J_{H} \: = \: \pt\otimes\eye_{\vec{x}}\otimes\eye_{\mrm{matter}} \: + \: \eye_{t}\otimes\eye_{\vec{x}}\otimes \ham \: \approx \: 0 \: ,
\end{equation}
\begin{equation}
\label{eq:QFT_JP}
\J_{\vec{P}} \: = \: \eye_{t}\otimes \hat{\vec{p}}\otimes\eye_{\mrm{matter}} \: - \: \eye_{t}\otimes\eye_{\vec{x}}\otimes \hat{P} \: \approx \: 0 \: ,
\end{equation}
where $\ham$ and $\hat{\vec{P}}$ are the Hamiltonian and momentum of the matter field (for example, a second quantized Klein-Gordon field), respectively. In a relativistically covariant theory, these generators commute and physical states would therefore be simultaneous eigenstates of the constraints. While interactions with a background field were not possible in the analysis of section \ref{sec:constraints} since they break time and/or space translation symmetry, we can treat interacting theories naturally in the context of the field theoretic generalization of Eq. (\ref{eq:QFT_HS}). This can be done by introducing field interactions directly at the level of the Hamiltonian $\ham \in \mathcal{L}(\hs_{\mathrm{matter}})$, for example a $\phi^{4}$-interacting field theory.
Another key aspect of relativistic quantum mechanics is the treatment of causality. In relativistic, first quantized systems, as we have discussed in the bulk of the paper, albeit using linear constraints, one encounters the standard quantum mechanical pitfall of acausal propagation. This is not surprising, since the constraint-based paradigm is designed so to reproduce results of standard relativistic quantum mechanics, where acausality is typically addressed using techniques of measurements, decoherence, etc. This indeed spells out the need for quantum field theory. In a quantum spacetime and constraint-based approach to quantum field theory using the Hilbert space structure of Eq. (\ref{eq:QFT_HS}), one can expect to work with both positive and negative frequency solutions and derive the main result of causality that operators commute at spacelike separations. This is outside the scope of the current paper and will be reported in future work focused on field theoretic constructions in such a quantum spacetime and constraint-based setup.
\\
\indent An interesting feature to note in the constraints we have discussed so far is the exclusive use of operators that do not couple different subfactors of $\hs$. They have a decoupled form \textit{i.e.} they are a collection of terms, each of which acts non-trivially only on a particular Hilbert space factor. Adding interaction terms that couple the spacetime Hilbert space $\hst\otimes\hsx$ to matter $\hs_{\mrm{matter}}$ could be useful in understanding effects like gravitational coupling and back-reaction. For example, interactions between the temporal degree of freedom and the system in the context of the Page-Wootters internal time (as described in section \ref{sec:time-space-spin-HS}) has been explored in Ref.  \cite{smith2019quantizing}. We saw in section \ref{sec:basis_changes} how different decompositions of the spacetime Hilbert space implemented by global basis changes can describe Lorentz transformations.
More generally, we can expect a broad class of unitary basis choices implementing different decompositions of Hilbert space to correspond to different choice of coordinate systems. 
The apparent freedom in the choice of decomposition of Hilbert space $\hs$ of Eq. (\ref{eq:PW_HS}) to choose a different clock/temporal degree of freedom $\hst$ and therefore different emergent dynamics for the system $\hs_S$ (in the context of the Page-Wootters formulation of section \ref{subsec:PW}) is often referred to as the ``Clock Ambiguity" \cite{Albrecht:1994bg,Albrecht:2007mm,Albrecht:2012wsd}. While the decoupled form of the constraints are rather special in their own right (such as Ref.  \cite{marletto2017evolution} where it is argued a decoupled form of the constraint can help ease the Clock Ambiguity) and some unitary transformations will preserve it (for instance, the global basis changes in section \ref{sec:basis_changes} to implement Lorentz transformations are such examples), not all unitary transformations will preserve this decoupled form. Investigating generic unitary basis changes for constraints containing interacting terms could shed light upon the nature of coordinates in the context of a quantum field theoretic setup for spacetime and matter and will be taken up in future work.  Not all decompositions may be allowed and some may be preferred over others in determining which degrees of freedom in $\hs$ make up the background spacetime and which make up the matter degrees of freedom. While definitely an interesting question that could have implications for the Hilbert space structure of quantum gravity, it is beyond the scope of this work and is left for future investigation. The interested reader is encouraged to look into a rich literature \cite{martinelli2019quantifying,hohn2018switch,vanrietvelde2018change,hohn2019switching, vanrietvelde2018switching} (and references therein) available on quantum frames of reference and its connections with Hamiltonian constraints.
\\
\indent We would also like to emphasize that while we have attempted to treat space and time on an equal footing in Hilbert space in the context of relativistic quantum mechanics, there are important and crucial differences between the nature of time and space. For instance, relativistic light cone structures demand causal influence in timelike directions and not spacelike, and the nature of time, more conventionally interpreted, seems to be inseparably intertwined with thermodynamics and the arrow of time \cite{Connes:1994hv,Rovelli:2009ee}. The interplay between space, time, and quantum mechanics can be better understood by re-examining crucial first principle ideas, which we believe to be an important direction of inquiry toward our understanding of quantum gravity. 

\section*{Disclosures and Acknowledgments}

On behalf of all authors, the corresponding author states that there is no conflict of interest. My manuscript has no associated data.
\\
\\
 I would like to thank Sean Carroll, Charles Cao, Aidan Chatwin-Davies, Swati Chaudhary and Frank Porter for helpful discussions during the course of this project. This material is based upon work supported by the U.S. Department of Energy, Office of Science, Office of High Energy Physics, under Award Number DE-SC0011632, as well as by the Walter Burke Institute for Theoretical Physics at Caltech and the Foundational Questions Institute. 

\bibliographystyle{utphys}
\bibliography{QM_spacetime_HS}

\end{document}